\documentclass[12pt,onecolumn]{article}

\usepackage{amsmath}
\usepackage{amssymb}
\usepackage{graphicx}
\usepackage{graphics}
\usepackage{epsfig}
\usepackage{array}
\usepackage[nomarkers,nofiglist,notablist,tablesfirst]{endfloat}

\begin{document}

\title{Adaptive Channel Allocation Spectrum Etiquette for Cognitive Radio
Networks}

\author{{Nie Nie   and   Cristina Comaniciu}\\
Department of Electrical and Computer Engineering\\
Stevens Institute of Technology,
Hoboken, NJ 07030\\
Email:\{nnie  ccomanic\}@stevens.edu
\date{}
\thanks{This work was supported in part by the NSF grant number: 527068}}
\maketitle

\begin{abstract}
In this work, we propose a game theoretic framework to analyze the
behavior of cognitive radios for distributed adaptive channel
allocation. We define two different objective functions for the
spectrum sharing games, which capture the utility of selfish users
and cooperative users, respectively. Based on the utility
definition for cooperative users, we show that the channel
allocation problem can be formulated as a potential game, and thus
converges to a deterministic channel allocation Nash equilibrium
point. Alternatively, a no-regret learning implementation is
proposed for both scenarios and it is shown to have similar
performance with the potential game when cooperation is enforced,
but with a higher variability across users. The no-regret learning
formulation is particularly useful to accommodate selfish users.
Non-cooperative learning games have the advantage of a very low
overhead for information exchange in the network.

We show that cooperation based spectrum sharing etiquette improves
the overall network performance at the expense of an increased
overhead required for information exchange.
\end{abstract}


\section{Introduction}

With the new paradigm shift in the FCC's spectrum management policy
\cite{FCC 2005} that creates opportunities for new, more
aggressive, spectrum reuse, cognitive radio technology lays the
foundation for the deployment of smart flexible networks that
cooperatively adapt to increase the overall network performance.
The cognitive radio terminology was coined by Mitola \cite{Mitola
2000}, and refers to a smart radio which has the ability to sense
the external environment, learn from the history, and make
intelligent decisions to adjust its transmission parameters
according to the current state of the environment.

The potential contributions of cognitive radios to spectrum sharing
and an initial framework for formal radio etiquette have been
discussed in \cite{Mitola 1999}. According to the proposed
etiquette, the users should listen to the environment, determine
the radio temperature of the channels and estimate their
interference contributions on their neighbors. Based on these
measurements, the users should react by changing their transmission
parameters if some other users may need to use the channel.

While it is clear that this etiquette promotes cooperation between
cognitive radios, the behavior of networks of cognitive radios
running distributed resource allocation algorithms is less well
understood.

As the cognitive radios are essentially autonomous agents that are
learning their environment and are optimizing their performance by
modifying their transmission parameters, their interactions can be
modeled using a game theoretic framework. In this framework, the
cognitive radios are the players and their actions are the
selection of new transmission parameters and new transmission
frequencies, etc., which influence their own performance, as well
as the performance of the neighboring players.

Game theory has been extensively applied in microeconomics, and
only more recently has received attention as a useful tool to
design and analyze distributed resource allocation algorithms (e.g.
\cite{Goodman_Mandayam 2001}-\cite{Menon_Reed 2004}). Some game
theoretic models for cognitive radio networks were presented in
\cite{Neel_Reed_Gilles 2002}, which has identified potential game
formulations for power control, call admission control and
interference avoidance in cognitive radio networks. The convergence
conditions for various game models in cognitive radio networks are
investigated in \cite{Neel_Reed_Gilles 2004}.

In this work, we propose a game theoretic formulation of the
adaptive channel allocation problem for cognitive radios. Our
current work assumes that the radios can measure the local
interference temperature on different frequencies and can adjust by
optimizing the information transmission rate for a given channel
quality (using adaptive channel coding) and by possibly switching
to a different frequency channel. The cognitive radios' decisions
are based on their perceived utility asociated with each possible
action. We propose two different utility definitions, which reflect
the amount of cooperation enforced by the spectrum sharing
etiquette. We then design adaptation protocols based on both a
potential game formulation, as well as no-regret learning
algorithms. We study the convergence properties of the proposed
adaptation algorithms, as well as the tradeoffs involved.

\section{System Model}

\begin{figure}[ht]
\centerline{ \epsfxsize=3.5 in\epsffile{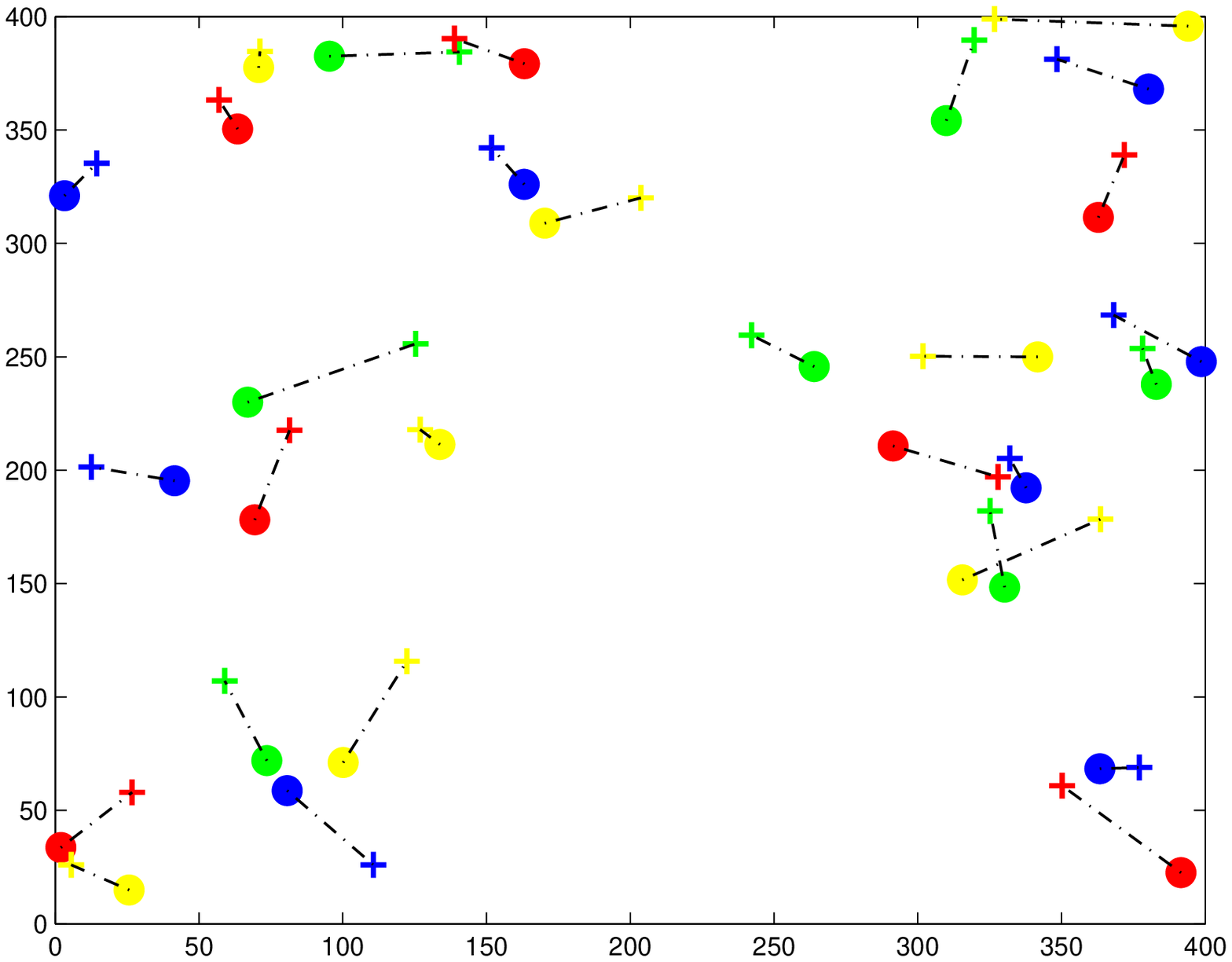}} \caption{A
snapshot of the nodes' positions and network topology}
\label{fig:pos}
\end{figure}

The cognitive radio network we consider consists of a set of $N$
transmitting-receiving pairs of nodes, uniformly distributed in a
square region of dimension $D^*\times D^*$. We assume that the
nodes are either fixed, or are moving slowly (slower than the
convergence time for the proposed algorithms). Fig. \ref{fig:pos}
shows an example of a network realization, where we used dashed
lines  to connect the transmitting node to its intended receiving
node.
The nodes measure the spectrum availability and decide on the
transmission channel. We assume that there are $K$ frequency
channels available for transmission, with $K < N$. By
distributively selecting a transmitting frequency, the radios
effectively construct a channel reuse distribution map with reduced
co-channel interference.

The transmission link quality can be characterized by a required
Bit Error Rate target (BER), which is specific for the given
application. An equivalent SIR target requirement can be
determined, based on the modulation type and the amonunt of channel
coding.

The Signal-to-Interference Ratio (SIR) measured at the receiver $j$
associated with transmitter $i$ can be expressed as:
\begin{equation}
SIR_{ij}=\frac{p_{i}G_{ij}}{\sum^{N}_{k=1, k\neq i}p_{k}G_{kj}I(k,
j)}, \label{eq:sir}
\end{equation}
where $p_i$ is the transmission power at transmitter $i$, $G_{ij}$
is the link gain between transmitter $i$ and receiver $j$. $I(i,j)$
is the interference function characterizing the interference
created by node $i$ to node $j$ and is defined as
\begin{equation}
I(i,j)=\left\{ \begin{array}{ll} 1 & \textrm{if transmitters $i$
and
$j$ are transmitting}\\
   & \textrm{ over the same channel }\\
0 & \textrm{otherwise}\\
\end{array}\right.
\end{equation}

Analyzing (\ref{eq:sir}) we see that in order to maintain a certain
BER constraint the nodes can adjust at both the physical and the
network layer level. At the network level, the nodes can minimize
the interference by appropriately selecting the transmission
channel frequency. At the physical layer, power control can reduce
interference and, for a feasible system,  results in all users
meeting their SIR constraints. Alternatively, the target SIR
requirements can be changed (reduced or increased) by using
different modulation levels and various channel coding rates. As an
example of adaptation at the physical layer, we have assumed that
for a
 fixed transmission power level, software defined radios enable the
 nodes to adjust their transmission rates and consequently the required
 SIR targets by varying the amount of channel coding for a data packet.

For our simulations we have assumed that all users have packets to
transmit at all times (worst case scenario). Multiple users are
allowed to transmit at the same time over a shared channel. We
assume that users in the network are identical, which means they
have an identical action set and identical utility functions
associated with the possible actions.

The BER requirement selected for simulations is $10^{-3}$, and we
assume the use of a Reed-Muller channel code RM $(1,m)$. In table
\ref{tb:coderates} we show the coding rate combinations and the
corresponding SIR target requirements used for our simulations
\cite{Hasan}.
\begin{table}
\caption{Code rates of Reed-Muller code RM $(1,m)$ and
corresponding SIR requirement for target BER=$10^{-3}$}
\label{tb:coderates}
\begin{center}
\begin{tabular}{|c|l|r|}
  \hline
  m & Code Rate & SIR (dB)\\
  \hline
  2 & 0.75 & 6 \\
  3 & 0.5 & 5.15 \\
  4 & 0.3125 & 4.6 \\
  5 & 0.1875 & 4.1 \\
  6 & 0.1094 & 3.75 \\
  7 & 0.0625 & 3.45 \\
  8 & 0.0352 & 3.2 \\
  9 & 0.0195 & 3.1 \\
  10 & 0.0107 & 2.8 \\
  \hline
\end{tabular}
\end{center}
\end{table}
\section{A Game Theoretic Framework}

Game theory represents a set of mathematical tools developed for
the purpose of analyzing the interactions in decision processes.
Particularly, we can model our channel allocation problem as the
outcome of a game, in which the players are the cognitive radios,
their actions (strategies), are the choice of a transmitting
channel and their preferences are associated with the quality of
the channels. The quality of channels is determined by the
cognitive radios by measurements on different radio frequencies.

We model our channel allocation problem as a normal form game,
which can be mathematically defined as $\Gamma=\{
N,\{S_i\}_{i\in{N}},\{U_i \}_{i\in{N}} \}$, where  N is the finite
set of players (decision makers), and $S_i$ is the set of
strategies associated with player $i$. We define $\mathbb{S}=\times
S_i, i\in{N}$ as the strategy space, and $U_i$:
$\mathbb{S}\rightarrow \mathbb{R}$ as the set of utility functions
that the players associate with their strategies. For every player
$i$ in game $\Gamma$, the utility function, $U_i$, is a function of
$s_i$, the strategy selected by player $i$, and of the current
strategy profile of its opponents: $s_{-i}$.

In analyzing the outcome of the game, as the players make decisions
independently and are influenced by the other players' decisions,
we are interested to determine if there exist a convergence point
for the adaptive channel selection algorithm, from which no player
would deviate anymore, i.e. a Nash equilibrium (NE). A strategy
profile for the players, $S=[s_1, s_2, ... , s_N ]$, is a NE if and
only if
\begin{equation}
U_i(S)\geq U_i(s^{'}_i, s_{-i}), \  \forall i\in{N}, s'_i\in{S_i}.
\label{eq:NE}\end{equation} If the equilibrium strategy profile in
(\ref{eq:NE}) is deterministic, a pure strategy Nash equilibrium
exists. For finite games, even if a pure strategy Nash equilibrium
does not exist, a mixed strategy Nash equilibrium can be found
(equilibrium is characterized by a set of probabilities assigned to
the pure strategies).

As becomes apparent from the above discussion, the performance of
the adaptation algorithm depends significantly on the choice of the
utility function which characterizes the preference of a user for a
particular channel. The choice of a utility function is not unique.
It must be selected to have physical meaning for the particular
application, and also to have appealing mathematical properties
that will guarantee equilibrium convergence for the adaptation
algorithm. We have studied and proposed two different utility
functions, that capture the channel quality, as well as the level
of cooperation and fairness in sharing the network resources.

\subsection{Utility Functions}

The first utility function (U1) we propose accounts for the case of
a ``selfish'' user, which values a channel based on the level of
interference perceived on that particular channel:
\begin{equation}
U1_i(s_i, s_{-i})=-\sum^{N}_{j\neq{i}, j=1}p_jG_{ij}f(s_j, s_i).
\label{eq:U1}
\end{equation}
\hspace*{+5cm}$\forall i=1, 2, ..., N$

For the above definition, we denoted P=[$p_1$,$p_2$,...,$p_N$] as
the transmission  powers for the N radios,
S=[$s_1$,$s_2$,...,$s_N$] as the strategy profile and $f(s_i, s_j)$
as an interference function:
\begin{equation}
f(s_i, s_j)=\left\{ \begin{array}{ll} 1 & \textrm{if $s_j=s_i$,
transmitter $j$ and $i$ choose}\\
& \textrm{the same strategy (same channel)}\\
0 & \textrm{otherwise}\\
\end{array}\right.
\nonumber
\end{equation}

This choice of the utility function requires a minimal amount of
information for the adaptation algorithm, namely the interference
measurement of a particular user on different channels.

The second utility function we propose accounts for the
interference seen by a user on a particular channel, as well as for
the interference this particular choice will create to neighboring
nodes. Mathematically we can define U2 as:

\begin{equation}
\mbox{\hspace*{-5cm}} U2_i(s_i, s_{-i})= \nonumber
\end{equation}
\vspace*{-.5cm}
\begin{equation}
 -\sum^{N}_{j\neq{i},
j=1}p_jG_{ij}f(s_j, s_i) -\sum^{N}_{j\neq{i}, j=1}p_iG_{ji}f(s_i,
s_j) \label{eq:U2}
\end{equation}
\hspace*{+5cm}$\forall i=1, 2, ..., N$

The complexity of the algorithm implementation will increase for
this particular case, as the algorithm will require probing packets
on a common access channel for measuring and estimating the
interference a user will create to neighboring radios.
%Finally, we consider to further increase the fairness of the
%channel allocation algorithm by diminishing the utility gain a
%user gets by further increasing the achieved SIR beyond a
%specified threshold $b$. The reason behind this choice is that
%this function might put more weight on other users' performance,
%when the current user's performance is already reasonably high. We
%thus define the utility function U3 as:
%
%\begin{equation}
%%U_i(s_i, s_{-i})=\log_{b}{\sum^{N}_{j\neq{i}, j=1}p_jG_{ij}f(s_j,
%%s_i)} -\sum^{N}_{j\neq{i}, j=1}p_iG_{ji}f(s_i, s_j)
%U_i(s_i, s_{-i})=\left\{ \begin{array}{ll} - \left(\frac{p_iG_{ii}}{b}\right) -\sum^{N}_{j\neq{i}, j=1}p_iG_{ji}f(s_i, s_j) & \textrm{if $SIR_i \geq b$}\\
%
%-\sum^{N}_{j\neq{i}, j=1}p_jG_{ij}f(s_j, s_i)
%-\sum^{N}_{j\neq{i}, j=1}p_iG_{ji}f(s_i, s_j) & \textrm{otherwise}\\
%\end{array}\right.
%\label{eq:U3}
%\end{equation}
%\hspace*{+5cm}$\forall i=1, 2, ..., N$
%where $SIR_i$ is the achieved SIR at receiver $i$.

The above defined utility functions, characterize a user's level of
cooperation and support a selfish and a cooperative spectrum
sharing etiquette, respectively.

\subsection{A Potential Game Formulation}

In the previous section we have discussed the choice of the utility
functions based on the physical meaning criterion. However, in
order to have good convergence properties for the adaptation
algorithm we need to impose some mathematical properties on these
functions. There are certain classes of games that have been shown
to converge to a Nash equilibrium when a best response adaptive
strategy is employed. In what follows, we show that for the $U2$
utility function, we can formulate an exact potential game, which
converges to a pure strategy Nash equilibrium solution.

Characteristic for a potential game is the existence of a potential
function that exactly reflects any unilateral change in the utility
function of any player. The potential function models the
information associated with the improvement paths of a game instead
of the exact utility of the game \cite{Monder 1996}.

An exact potential function is defined as a function
\begin{equation}
\mbox{\textbf{P}}: \mathbb{S}\rightarrow  \mathbb{R}, \ \mbox{if
for all } i, \mbox{and } s_i, \ s'_i \in {S_i}, \nonumber
\end{equation}
\noindent with the property that
\begin{equation}
U_i(s_i, s_{-i})-U_i(s'_i,s_{-i}) = P(s_i,s_{-i})-P(s'_i,s_{-i}).
\label{eq:prop}
\end{equation}
If a potential function can be defined for a game, the game is an
exact potential game. In an exact potential game, for a change in
actions of a single player the change in the potential function is
equal to the value of the improvement deviation. Any potential game
in which players take actions sequentially converges to a pure
strategy Nash equilibrium that maximizes the potential function.

For our previously formulated channel allocation game with utility
function $U2$, we can define an exact potential function to be
\hspace*{-2cm}
\begin{equation} Pot(S)=Pot(s_i, s_{-i})= \nonumber
\end{equation}
\begin{equation}
\sum^{N}_{i=1}\left( -\frac{1}{2} \sum^{N}_{j\neq{i},
j=1}p_jG_{ij}f(s_j, s_i) -\frac{1}{2}\sum^{N}_{j\neq{i},
j=1}p_iG_{ji}f(s_i, s_j) \right) \label{eq:Pot}
\end{equation}
\hspace*{+5cm}$\forall i=1, 2, ..., N$.\\
The function in (\ref{eq:Pot}) essentially reflects the network
utility. It can be seen thus that the potential game property
(\ref{eq:prop}) ensures that an increase in individual users'
utilities contributes to the increase of the overall network
utility. We note that this property holds only if users take
actions sequentially, following a best response strategy.

The proof that equation(\ref{eq:Pot}) is an exact potential
function is given in the Appendix.

%\%hspace{-.4cm} \textsc{lemma 1}  \textit{\\
%%\hspace*{+0.5cm}\textit{Proof:}  Shown in Appendix 1.

Consequently, to ensure convergence for the spectrum allocation
game, either a centralized or a distributed scheduler should be
deployed. In an ad hoc network, the latter solution is preferable.
To this end, we propose a random access for decision making in
which each user is successful with probability $p_a=1/N$. More
specifically, at the begining of each time slot, each user flips a
coin with probability $p_a$, and, if successful, makes a new
decision based on the current values for the utility functions for
each channel; otherwise takes no new action. We note that the
number of users that attempt to share each channel, can be
determined from channel listening as we will detail shortly. The
proposed random access ensures that on average exactly one user
makes decisions at a time, but of course has a nonzero probability
to have two or more users taking actions simultaneously. We have
determined experimentally that the convergence of the game is
robust to this phenomenon: when two or more users simultaneously
choose channels, the potential function may temporarily decrease
(decreasing the overall network performance) but then the upward
monotonic trend is re-established.

The proposed potential game formulation requires that users should
be able to evaluate the candidate channels' utility function $U2$.
To provide all the information necessary to determine $U2$, we
propose a signaling protocol based on a three way handshake
protocol. The signaling protocol is somewhat similar to the RTS-CTS
packet exchange for the IEEE 802.11 protocol, but intended as a
call admission reservation protocol, rather than packet access
reservation protocol. When a user needs to make a decision on
selecting the best transmission frequency (a new call is initiated
or terminated, and user is successful in the Bernoulli trial),
 such a handshaking is initiated. In contrast to the
RTS-CTS reservation mechanism, the signaling packets, START,
START\_CH, ACK\_START\_CH (END, ACK\_END) in our protocol, are not
used for deferring transmission for the colliding users, but rather
to measure the interference components of the utility functions for
different frequencies and to assist in computing the utility
function. The signaling packets have a double role: to announce the
action of the current user to select a particular channel for
transmission, and to serve as probing packets for interference
measurements on the selected channel. The signaling packets are
transmitted with a fixed transmission power on a common control
channel. To simplify the analysis, we assume that no collisions
occur on the common control channel. As we mentioned before, the
convergence of the adaptation algorithm was experimentally shown to
be robust to collision situations. For a better frequency planning,
it is desirable to use a higher transmission power for the
signaling packets than for the transmitted packets. This will
permit the users to learn the potential interferers over a larger
area. For our simulations, we have selected the ratio of
transmitted powers between signaling and data packets to be equal
to 2.

We note that the $U2$ utility function has two parts: a) a measure
of the interference created by others on the desired user $I_d$; b)
a measure of the interference created by the user on its neighbors'
transmissions $I_o$. The first part of $U2$ can be estimated at the
receiving node, while the second part can only be estimated at the
transmitter node. Therefore, the protocol requires that both
transmitter and receiver listen to the control channel, and each
maintain an information table on all frequencies, similar to the
NAV table in 802.11. In what follows, we outline the steps of the
protocol.

\vspace{0.5cm} {\em Protocol steps:}

\begin{enumerate}
\item Bernoulli trial with $p_a$\\
\hspace{2cm} if 0, listen to the common control channel; {\em break}.\\
\hspace{2cm} if 1, go to 2) \item Transmitter sends START packet:
includes current estimates for the interference created to
neighboring users on all possible frequencies, $I_o(f)$ (this
information is computed based on information saved in the Channel
Status Table); \item Receiver computes current interference
estimate for the user $I_d(f)$, determines $U2(f) = I_d(f) +
I_o(f)$ for all channels, and decides on the channel with the
highest $U2$ (in case of equality, the selection is randomized,
with equal probability of selecting the channels); \item Receiver
includes the newly selected channel information on a signaling
packet $START\_CH$ which is transmitted on the common channel;
\item Transmitter sends $ACK\_START\_CH$ which acknowledges the
decision of transmitting on the newly selected frequency, and
starts transmitting on the newly selected channel; \item All the
other users (transmitters and receivers) that heard the $START\_CH$
and $ACK\_START\_CH$ packets update their Channel Status Tables
(CST) accordingly.
\end{enumerate}
\vspace{0.5cm}

We note that when a call ends, only a two-way handshake is
required: END, ACK\_END to announce the release of the channel for
that particular user. Upon hearing these end-of-call signaling
packets, all transmitters and receivers,  update their CSTs
accordingly.

We can see that a different copy of the CST should be kept at both
the transmitter and the receivers (CST\_t and CST\_r,
respectively). The entries of each table will contain the
neighboring users that have requested a channel, the channel
frequency, and the estimated link gain to the transmitter/receiver
of that particular user (for CST\_r and CST\_t, respectively).

%where CHL denotes
%the options of data transmission channel; $RX\_S$ is the received
%signal power at the intended receiver; $RX\_I$ is the received
%interference power at the intended receiver; $I\_to_1$ is the
%interference caused by this user at receiver $1$, similarly for
%$I\_to\_2$ ... $I\_to\_N$; The column of $I\_to\_i$ for user $i$
%is set to $0$ following a reasonable line.

The proposed potential game framework has the advantage that an
equilibrium is reached very fast following a best response dynamic,
but requires substantial information on the interference created to
other users and additional coordination for sequential updates. We
note however, that the sequential updates procedure also resolves
the potential conflicts on accessing the common control channel.

The potential game formulation is suitable for designing a
cooperative spectrum sharing ettiquette, but cannot be used to
analyze scenarios involving selfish users, or scenarios involving
heterogeneous users (with various utility functions corresponding
to different QoS requirements). In the following section, we
present a more general design approach, based on no-regret learning
techniques, which alleviates the above mentioned problems.

\subsection{$\Phi$-No-Regret Learning for Dynamic Channel Allocation}

While we showed in the previous section that the game with the $U2$
utility function fits the framework of an exact potential game, the
U1 function lacks the necessary symmetry properties that will
ensure the existence of a potential function. In order to analyze
the behavior of the selfish users game, we resort to the
implementation of adaptation protocols using regret minimization
learning algorithms. No regret learning algorithms are
probabilistic learning strategies that specify that players explore
the space of actions by playing all actions with some non-zero
probability, and exploit successful strategies by increasing their
selection probability. While traditionally, these types of learning
algorithms have been characterized using a regret measure (e.g.
external regret is defined as the difference between the payoffs
achieved by the strategies prescribed by the given algorithm, and
the payoffs obtained by playing any other fixed sequence of
decisions in the worst case), more recently, their performance have
been related to game theoretic equilibria.

A general class of no-regret learning algorithms called
$\Phi$-no-regret learning algorithm are shown in \cite{Greenwald
2003} to relate to a class of equilibria named $\Phi$-equilibria.
No-external-regret and no-internal regret learning algorithms are
specific cases of $\Phi$-no-regret learning algorithm. $\Phi$
describes the set of strategies to which the play of a learning
algorithm is compared. A learning algorithm is said to be
$\Phi$-no-regret if and only if no regret is experienced for
playing as the algorithm prescribes, instead of playing according
to any of the transformations of the algorithm's play prescribed by
elements of $\Phi$. It is shown in \cite{Greenwald 2003} that the
empirical distribution of play of $\Phi$-no-regret algorithms
converges to a set of $\Phi$-equilibria. It is also shown that
no-regret learning algorithms have the potential to learn mixed
strategy (probabilistic) equilibria. We note that Nash equilibrium
is not a necessary outcome of any $\Phi$-no regret learning
algorithm \cite{Greenwald 2003}.

We propose an alternate solution for our spectrum sharing problem,
based on a no-external-regret learning algorithm with exponential
updates, proposed in \cite{Freund 1995}.

Let $U^{t}_{i}(s_i)$ denote the cumulative utility obtained by user
$i$ through time $t$ by choosing strategy $s_i$:
$U^{t}_{i}(s_i)=\sum^{t}_{st=1}U_i(s_i, S^{st}_{-i}) $. For $\beta
> 0$, the weight (probability) assigned to strategy $s_i$ at time
$t+1$, is given by:
\begin{equation}\label{eq:weight}
w^{t+1}_{i}(s_i)=\frac{(1+\beta)^{U^{t}_{i}(s_i)} }{ \sum_{
s'_{i}\in S_i}(1+\beta)^{U^{t}_{i}(s'_i)}}.
\end{equation}

In \cite{Jafari 2001}, based on simulation results, it is shown
that the above learning algorithm converges to Nash equilibrium in
games for which pure strategy Nash equilibrium exists. We also show
by simulations that the proposed channel allocation no-regret
algorithm converges to a pure strategy Nash equilibrium for
cooperative users (utility $U2$), and to a mixed strategy
equilibrium for selfish users (utility $U1$).

By following our proposed learning adaptation process, the users
learn how to choose the frequency channels to maximize their
rewards through repeated play of the game.

For the case of selfish users, the amount of information required
by this spectrum sharing algorithm is minimal: users need to
measure the interference temperature at their intended receivers
(function $U1$) and to update their weights for channel selection
accordingly, to favor the channel with minimum interference
temperature (equal transmitted powers are assumed). We note that
the no-regret algorithm in (\ref{eq:weight}) requires that the
weights are updated for all possible strategies, including the ones
that were not currently played. The reward obtained if other
actions were played can be easily estimated by measuring the
interference temperature for all channels.

For the case of cooperative users, the information needed to
compute $U2$ is similar to the case of potential game formulation.
We note that, while the learning algorithm does not require
sequential updates to converge to an equilibrium, the amount of
information exchange on the common control channel requires
coordination to avoid collisions. One possible approach to reduce
the amount of signaling, would be to maintain the access scheme
proposed in the previous section, which would ensure that on
average only one user at the time will signal changes in channel
allocation.

\section{Simulation Results}

In this section, we present some numerical results to illustrate
the performance of the proposed channel allocation algorithms for
both cooperative and selfish users' scenarios. For simulation
purposes, we consider a  fixed wireless ad hoc network (as
described in the system model section) with $N=30$ and $D=200$ (30
transmitters and their receivers are randomly distributed over a
$200 m \times 200 m$ square area). The adaptation algorithms are
illustrated for a network of 30 transmitting radios, sharing $K=4$
available channels. A random channel assignment is selected as the
initial assignment and for a fair comparison, all the simulations
start from the same initial channel allocation.

\begin{figure}[ht]
\centerline{ \epsfxsize=3.5 in\epsffile{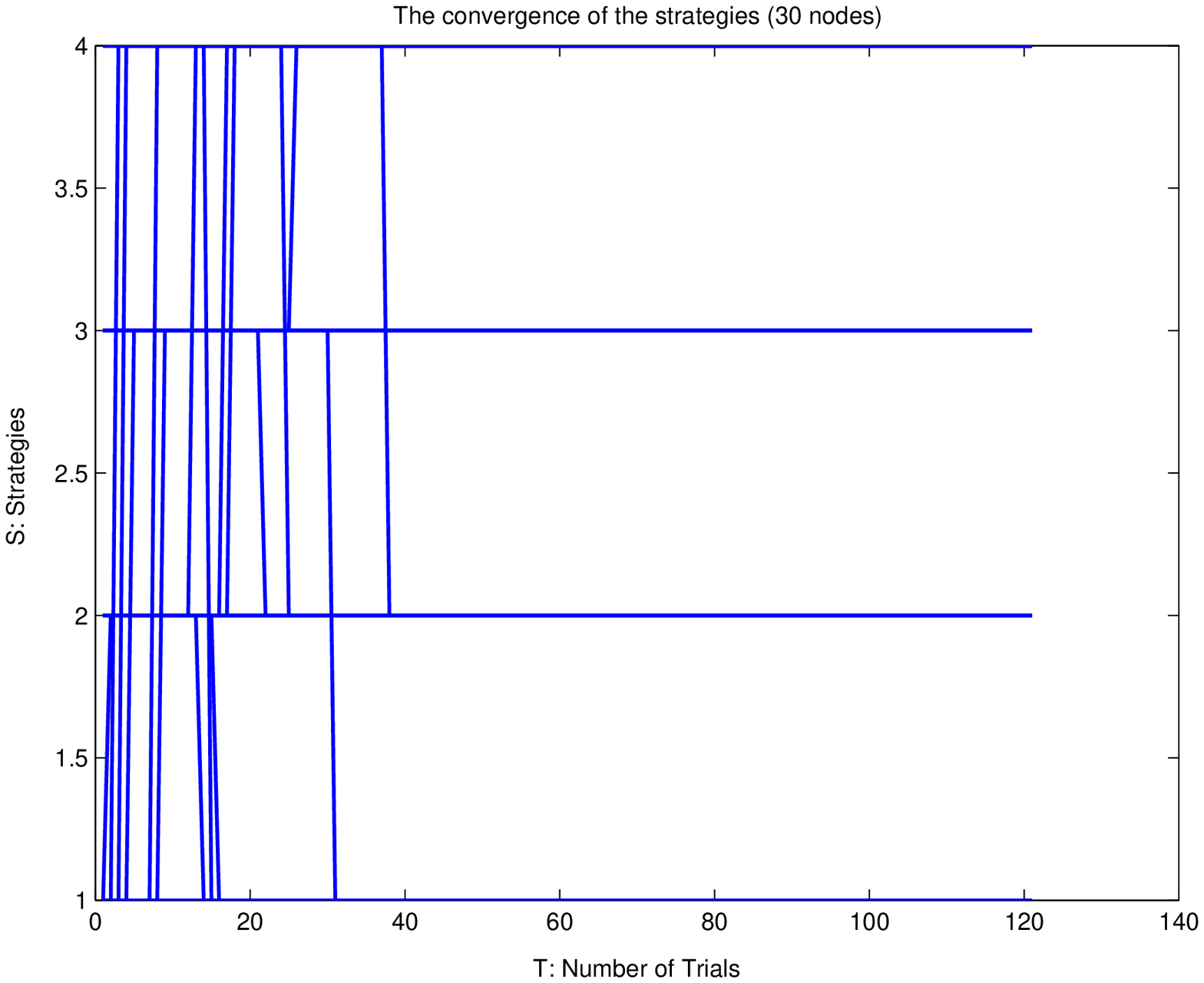}}
\caption{Potential game: convergence of users' strategies}
\label{fig:potconverge}
\end{figure}
\begin{figure}[ht]
\centerline{ \epsfxsize=3.5 in\epsffile{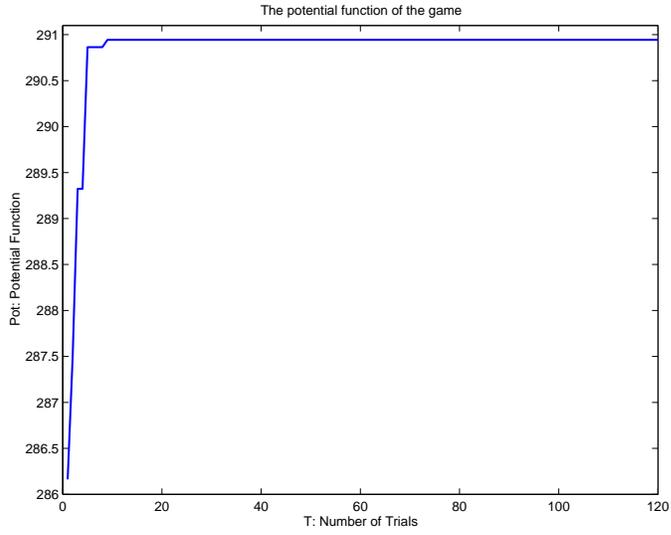}}
\caption{Evolution of potential function } \label{fig:potf}
\end{figure}
\begin{figure}[ht]
\centerline{ \epsfxsize=3.5 in\epsffile{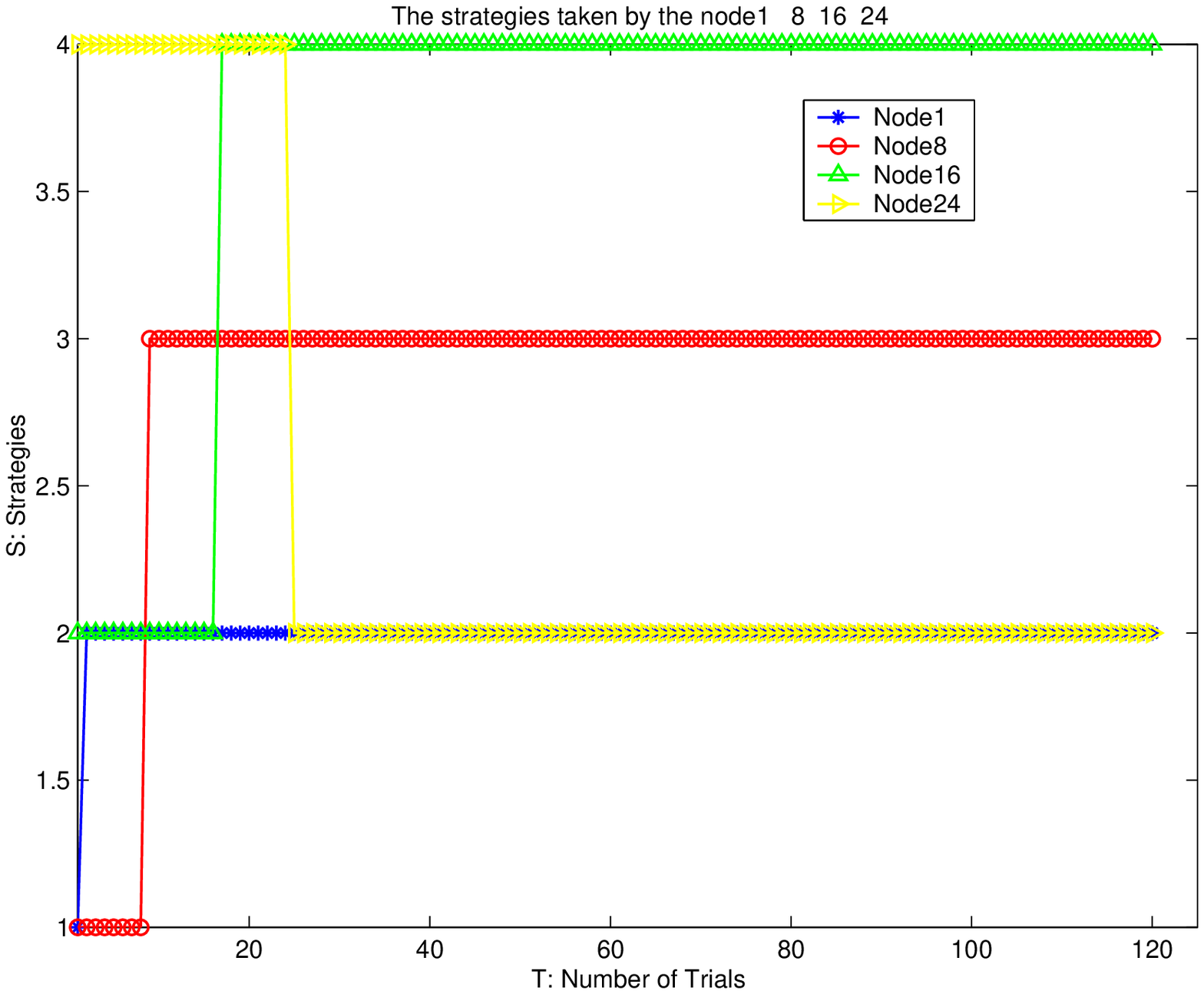}} \caption{
Potential game: strategy evolution for selected arbitrary users }
\label{fig:potactions}
\end{figure}
%\begin{figure}[ht]
%\centerline{ \epsfxsize=3.2 in\epsffile{potdemo2.eps}} \caption{
%Potential Game: Initial Channel Assignment and Final Channel
%Assignment with the throughput achieved at each users' receiver }
%\label{fig:potdemo2}
%\end{figure}

\begin{figure}[ht]
\centerline{ \epsfxsize=3.5 in\epsffile{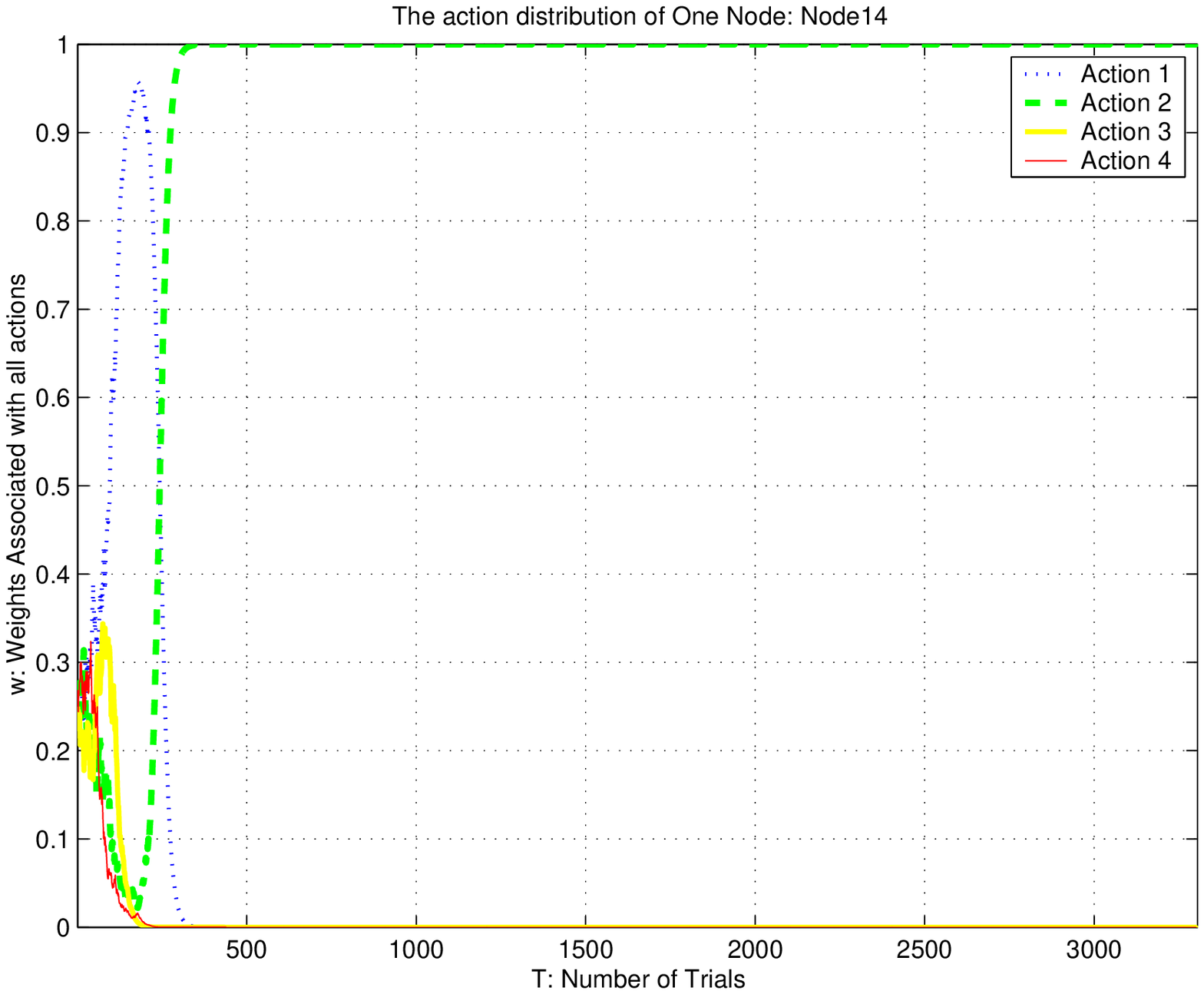}} \caption{
No-regret learning for cooperative users: weights distribution
evolution for an arbitrary user} \label{fig:fair1node}
\end{figure}

\begin{figure}[ht]
\centerline{ \epsfxsize=3.5 in\epsffile{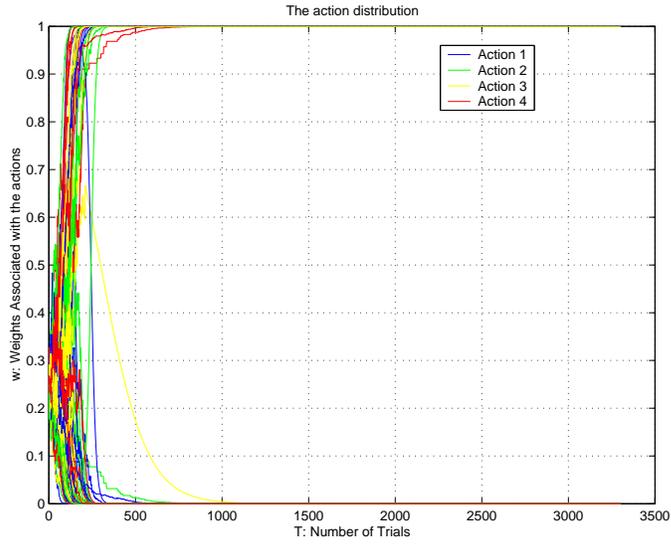}}
\caption{No-regret learning for cooperative users: weights
distribution evolution for all users} \label{fig:fairactions}
\end{figure}

We first illustrate the convergence properties of the proposed
spectrum sharing algorithms. We can see that for cooperative games,
both the potential game formulation, as well as the learning
solution converge to a pure strategy Nash equilibrium (Figures
\ref{fig:potconverge}, \ref{fig:potactions}, \ref{fig:fair1node}
and \ref{fig:fairactions}). In Figure \ref{fig:potf}, we illustrate
the changes in the potential function as the potential game
evolves, and it can be seen that indeed by distributively improving
their utility, the users positively affect the overall utility of
the network, which is approximated by the potential function.

\begin{figure}[ht]
\centerline{ \epsfxsize=3.5 in\epsffile{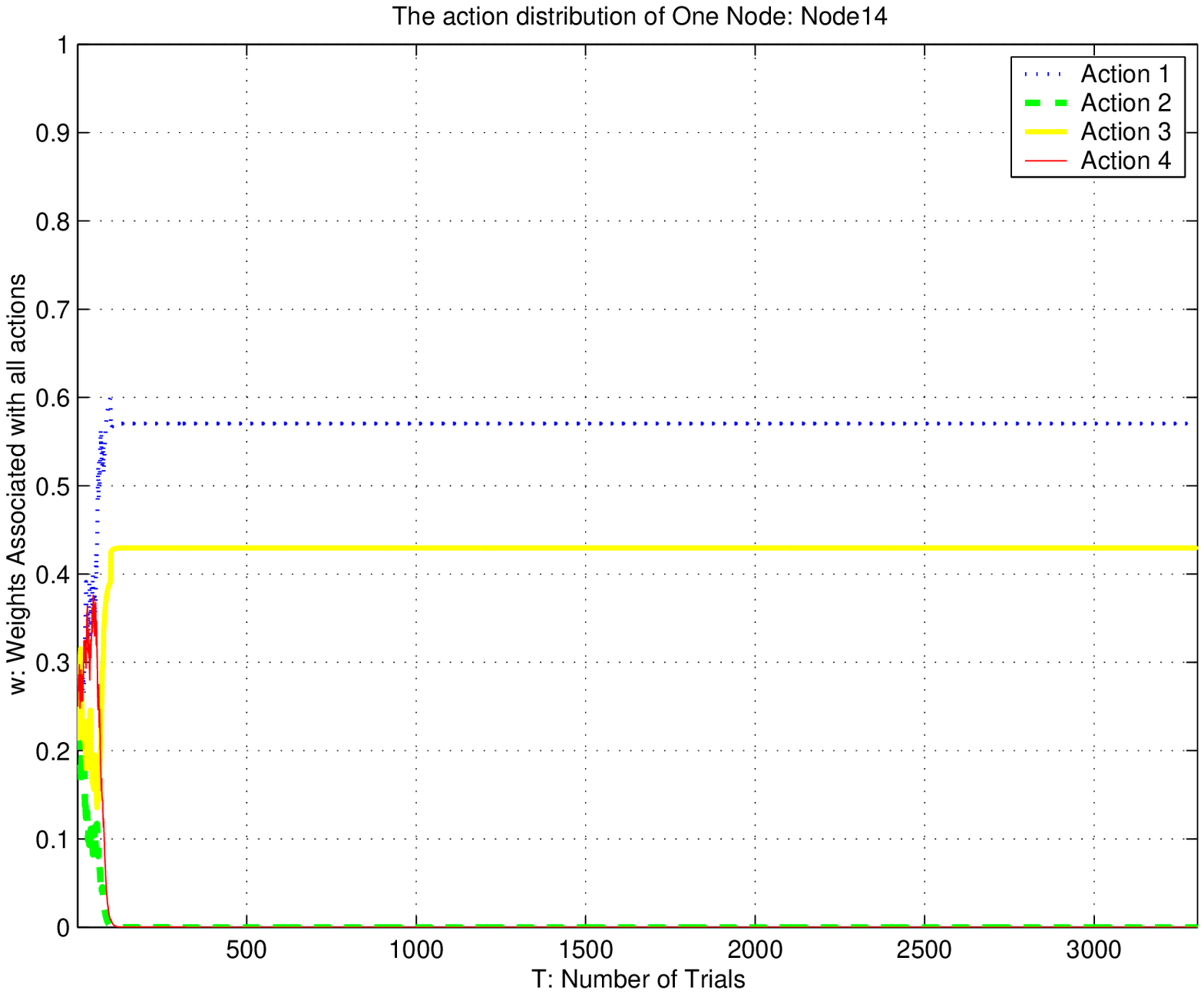}}
\caption{No-regret learning for selfish users: weights evolution for
an arbitrary user} \label{fig:selfnode14}
\end{figure}

\begin{figure}[ht]
\centerline{ \epsfxsize=3.5 in\epsffile{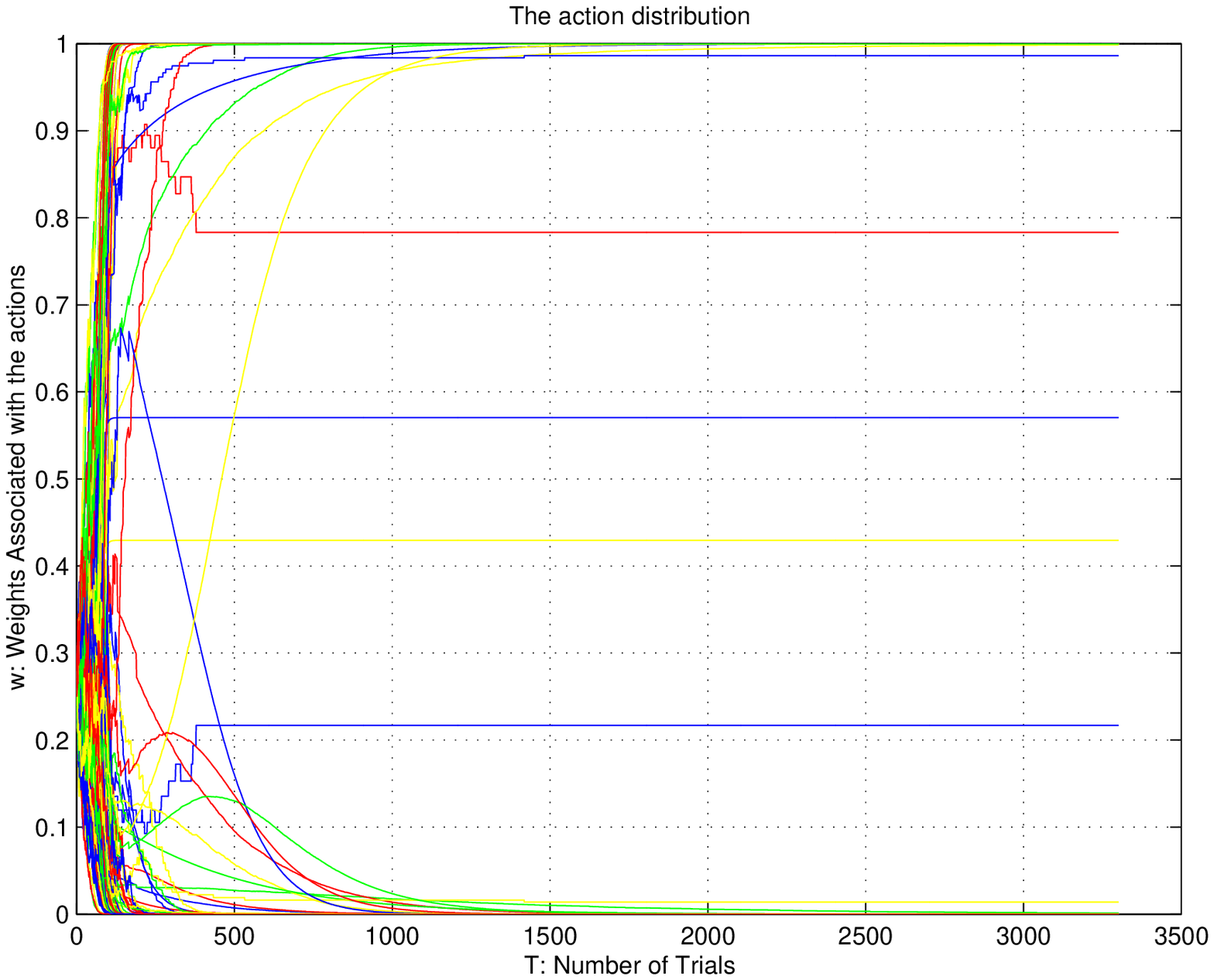}}
\caption{No-regret learning for selfish users: Evolution of weights
for all users} \label{fig:selfconverg}
\end{figure}

By contrast, the selfish users' learning strategy converges to a
mixed strategy equilibrium, as it can be seen in Figures
\ref{fig:selfnode14} and \ref{fig:selfconverg}.

As performance measures for the proposed algorithms we consider the
achieved SIRs and throughputs (adaptive coding is used to ensure a
certain BER target, as previously explained in Section II). We
consider the average performance per user as well as the
variability in the achieved performance (fairness), measured in
terms of variance and CDF.

We first give results for the potential game based algorithm.

\begin{figure}[ht]
\centerline{ \epsfxsize=3.5 in\epsffile{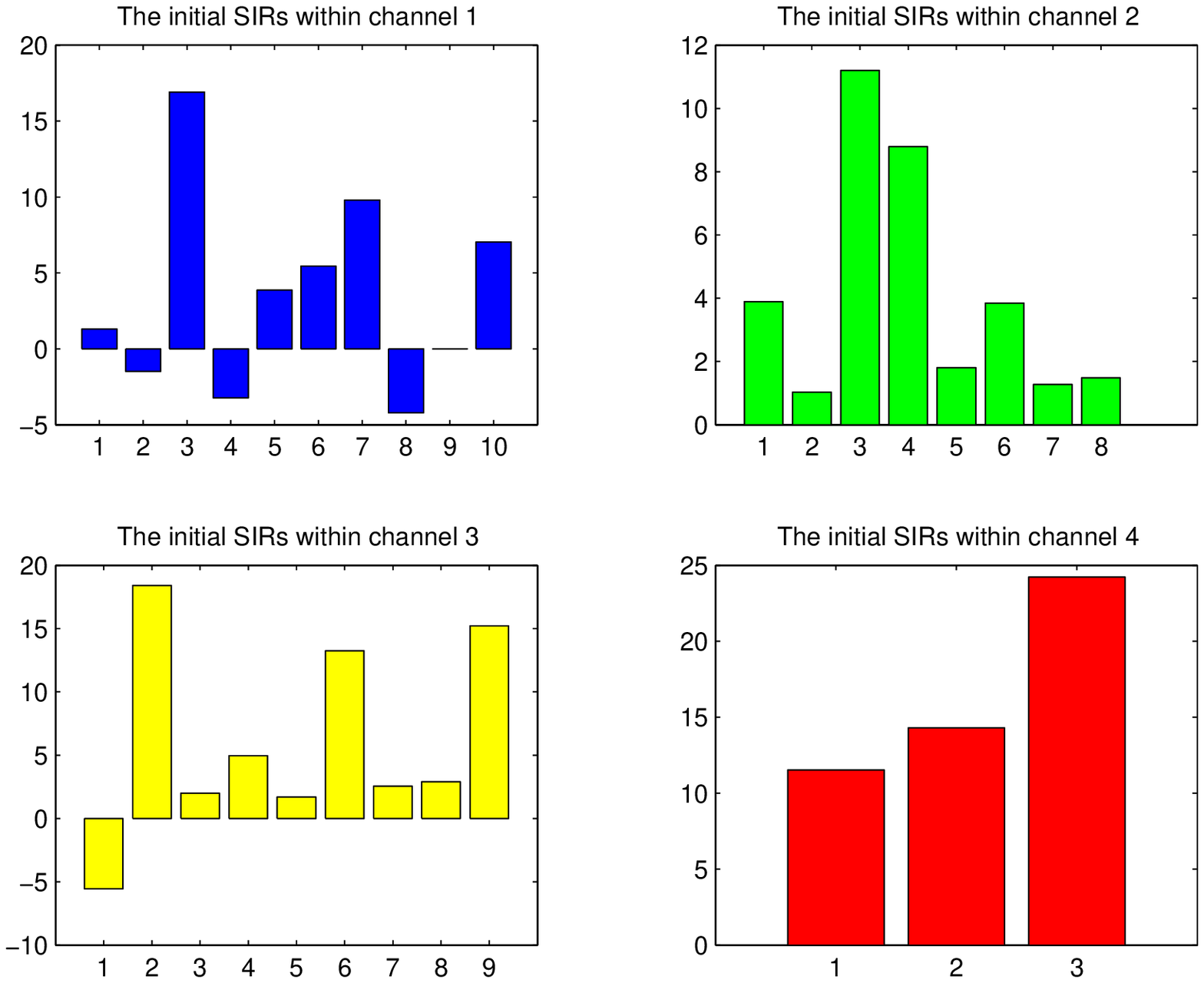}} \caption{
SIRs for initial channel assignment channels} \label{fig:potsirbar1}
\end{figure}
\begin{figure}[ht]
\centerline{ \epsfxsize=3.5 in\epsffile{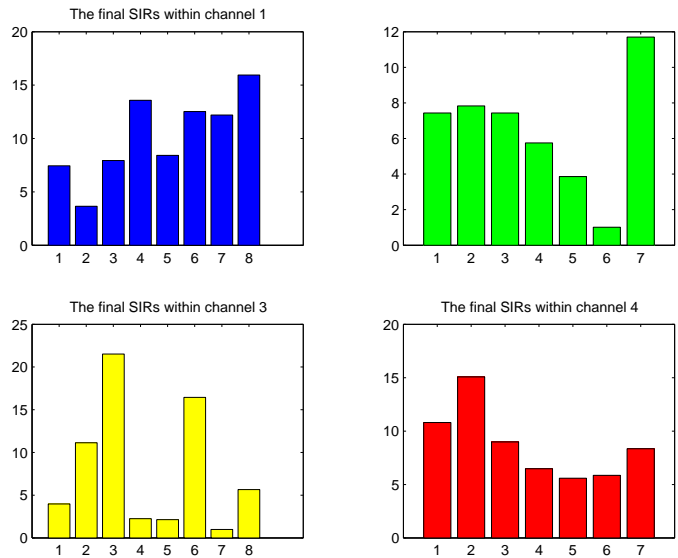}} \caption{
Potential Game: SIRs at final channel assignment}
\label{fig:potsirbar2}
\end{figure}
\begin{figure}[ht]
\centerline{ \epsfxsize=3.5 in\epsffile{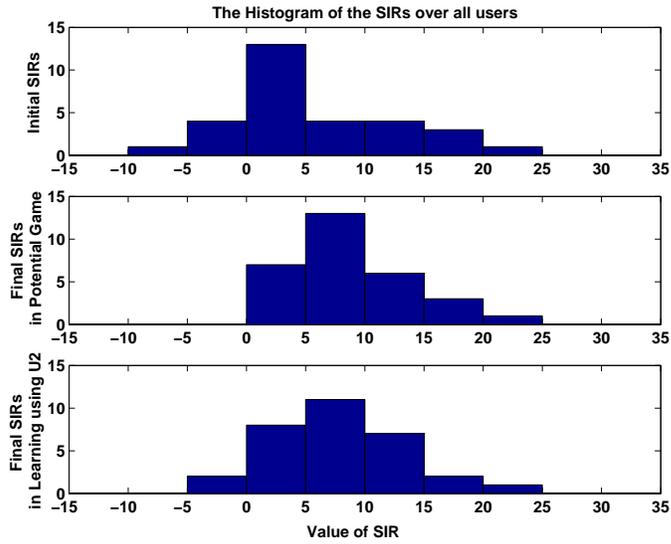}}
\caption{SIRs histogram. Initial Channel Assignment vs. Final
Channel Assignment} \label{fig:histsirdyspan}
\end{figure}
\begin{figure}[ht]
\centerline{ \epsfxsize=3.5 in\epsffile{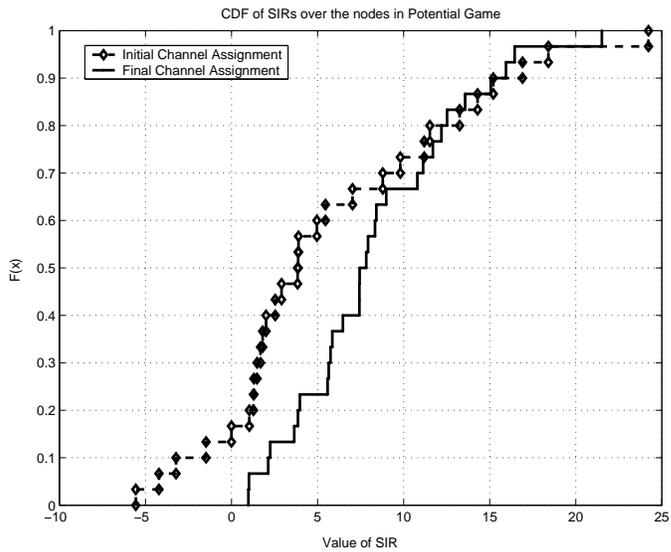}} \caption{CDF
for the achieved SIRs. Initial Channel Assignment vs. Final Channel
Assignment } \label{fig:potcdfsir}
\end{figure}
\begin{figure}[ht]
\centerline{ \epsfxsize=3.5 in\epsffile{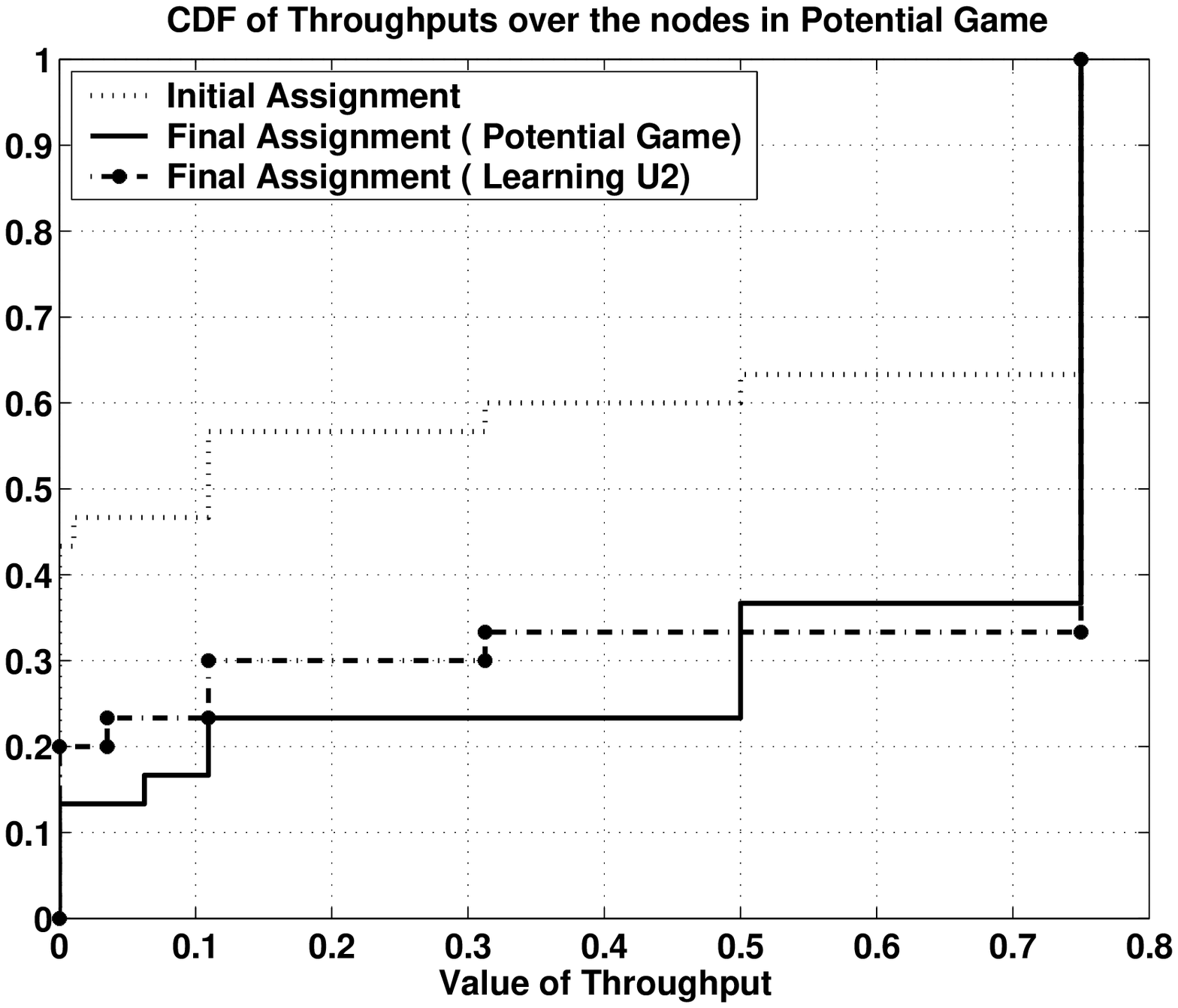}}
\caption{CDF for the achieved throughputs. Initial Channel
Assignment vs. Final Channel Assignment} \label{fig:cdfthdyspan}
\end{figure}
%\begin{figure}[ht]
%\centerline{ \epsfxsize=3.2 in\epsffile{pottotsirth.eps}}
%\caption{Total SIR and Total Throughput for the Initial and Final
%Channel Assignment} \label{fig:pottotsirth}
%\end{figure}
The choice of the utility function for this game enforces a certain
degree of fairness in distributing the network resources, as it can
be seen in figures \ref{fig:potsirbar1}, \ref{fig:potsirbar2},
\ref{fig:histsirdyspan}, and \ref{fig:potcdfsir}. Figures
\ref{fig:potsirbar1} and \ref{fig:potsirbar2} illustrate the SIR
achieved by the users on each of the 4 different channels for
initial and final assignments, respectively. An SIR improvement for
the users that initially had a low performance can be noticed, at
the expense of a slight penalty in performance for users with
initially high SIR. It can be seen in Figure
\ref{fig:histsirdyspan} that at the Nash equilibrium point, the
number of users having an SIR below 0 dB has been reduced.
Furthermore, figure \ref{fig:potcdfsir} shows that the percentage
of the users who have an SIR below 5 dB decreases from 60\% to
about 24\%, at the expense of a slight SIR decrease for users with
an SIR greater than 12.5 dB.

The advantage of the potential game is illustrated in figure
\ref{fig:cdfthdyspan}, in terms of the normalized achievable
throughput at each receiver.  For the initial channel assignment,
62\% of the users have a throughput less than 0.75. At the
equilibrium, this fraction is reduced to 38\%. Aggregate normalized
throughput improvements for the potential game formulation are
illustrated in Table \ref{tb:totsirth}.
\begin{table}
\caption{SIR and normalized throughput of all users at initial and
final channel assignment } \label{tb:totsirth}
\begin{center}
\begin{tabular}{|c|c|}
  \hline
    & Total Normalized Throughput \\
  \hline
  Initial & 9.4 \\
  Final (Potential Game) & 16.5 \\
  Final (Learning U2) & 15.3 \\
  \hline
\end{tabular}
\end{center}
\end{table}

\begin{figure}[ht]
\centerline{ \epsfxsize=3.5 in\epsffile{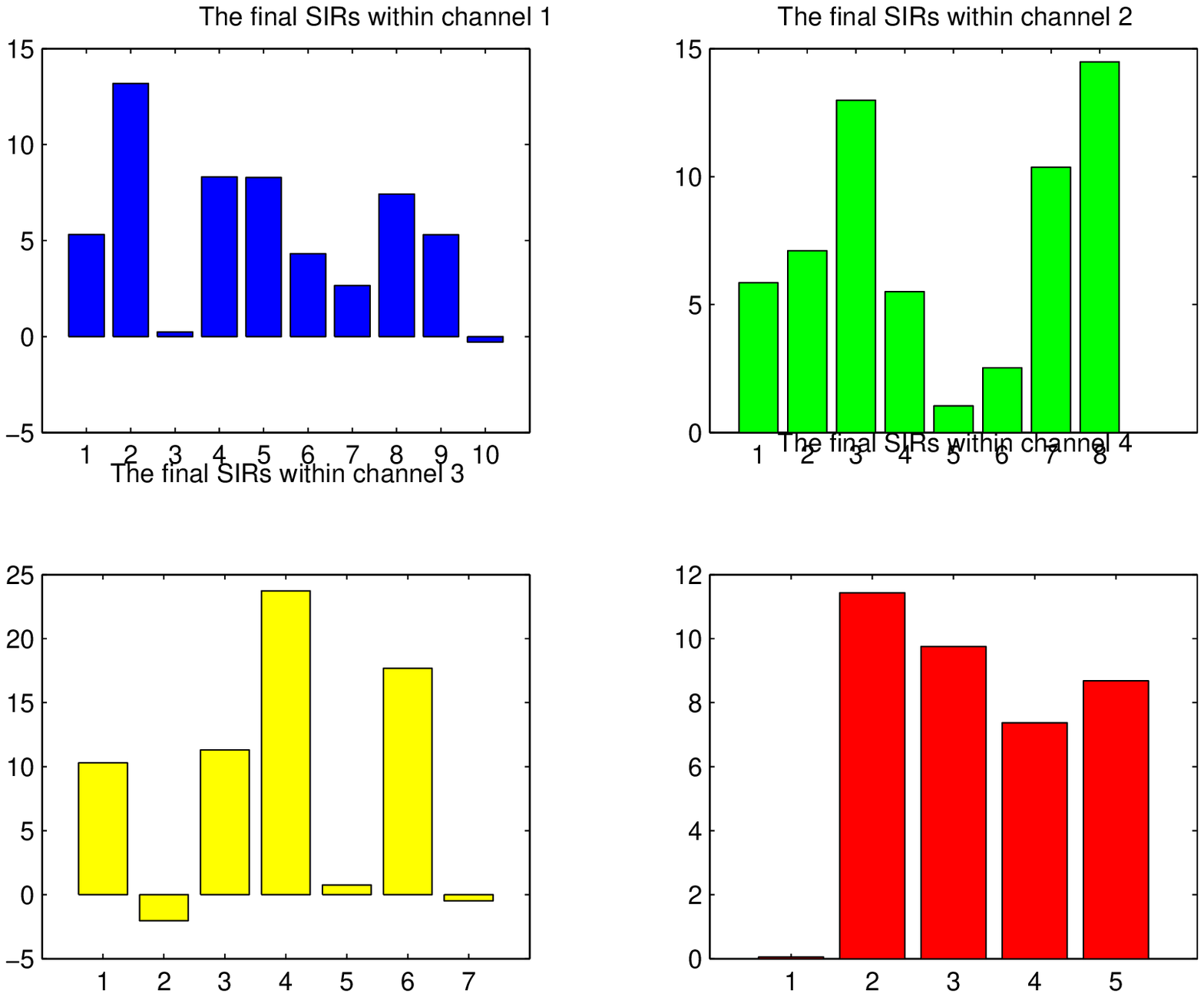}} \caption{
No-regret learning for cooperative users: SIR of users in different
channels at Nash equilibrium} \label{fig:u2sirbar2}
\end{figure}

Our simulation results show very similar performance for the
learning algorithm in cooperative scenarios, with the potential
game formulation. Figures \ref{fig:histsirdyspan} and
\ref{fig:u2sirbar2}, show the initial and final assignment for this
algorithm, as well as the achieved SIRs after convergence for all
users in the network. In terms of fairness, the learning algorithm
performs slighly worse than the potential game formulation (Figure
\ref{fig:cdfthdyspan}). However, even though the equilibrium point
for learning is different than that of the potential game, the two
algorithms achieve very close throughput performance (Table
\ref{tb:totsirth}).


As we previously mentioned, the learning algorithm for selfish
users does not lead to a pure strategy Nash equilibrium channel
allocation. In Figure \ref{fig:selfnode14} we illustrate the
convergence properties for an arbitrarily chosen user, which
converges to a  mixed strategy allocation: selects channel 1 with
probability 0.575 or channel 3 with probability 0.425. The
evolutions of the weights for all the users in the network are
shown in Figure \ref{fig:selfconverg}.


We compare the performance of the proposed algorithms for both
cooperative and non-cooperative scenarios. The performance measures
considered are the average SIR, average throughput per user, and
total average throughput for the network. At the beginning of each
time slot, every user will either choose the same equilibrium
channel for transmission (in cooperative games with pure strategy
Nash equilibrium solutions), or will choose a channel to transmit
with some probability given by the mixed strategy equilibrium (i.e.
learning using U1). In the random channel allocation scheme, every
user chooses a channel with equal probability from a pool of four
channels.

\begin{figure} [ht]
\centerline{ \epsfxsize=3.5 in\epsffile{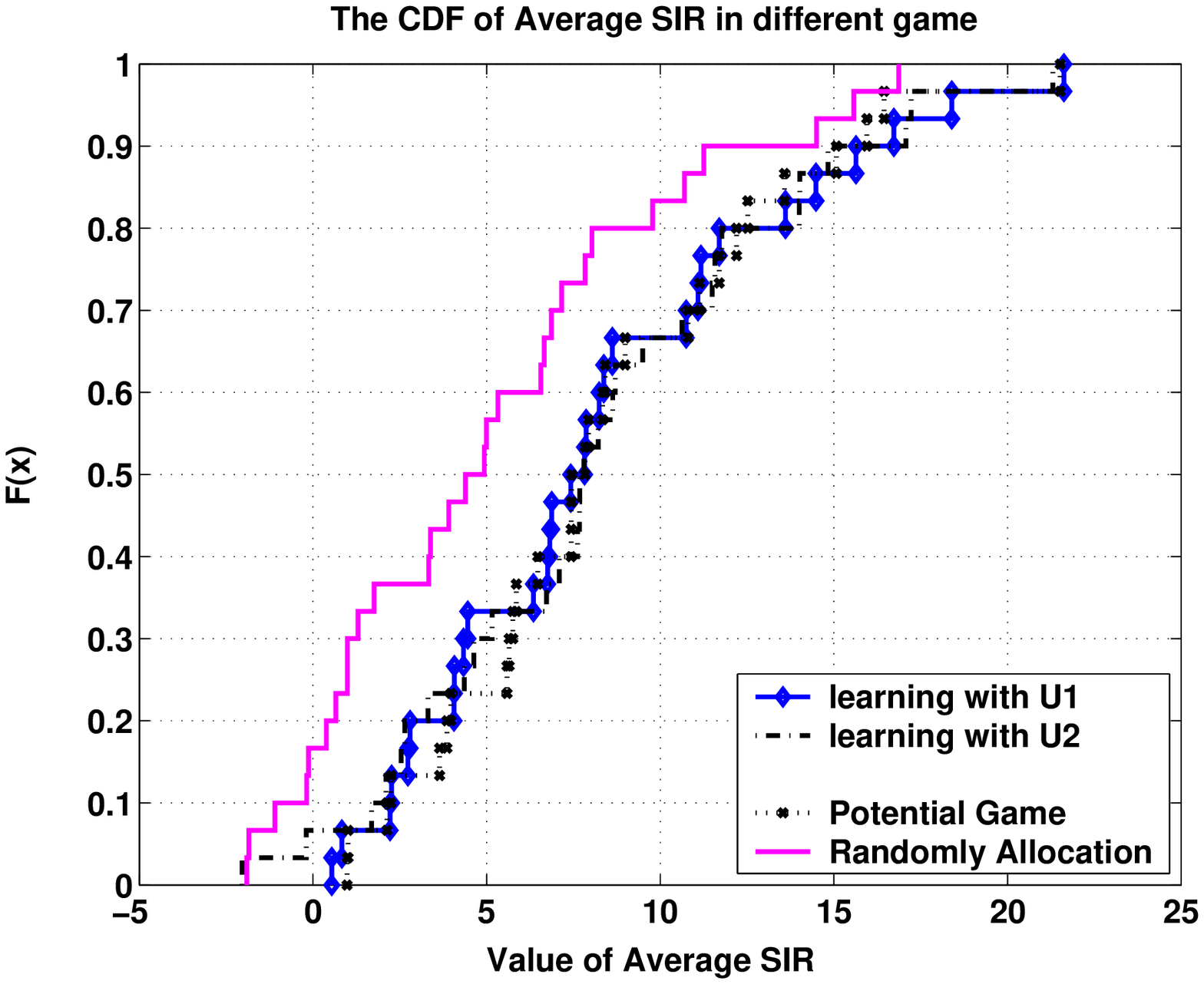}}
\caption{The CDF of Time Average SIRs} \label{fig:avgcdfsirdyspan}
\end{figure}

\begin{figure} [ht]
\centerline{ \epsfxsize=3.5 in\epsffile{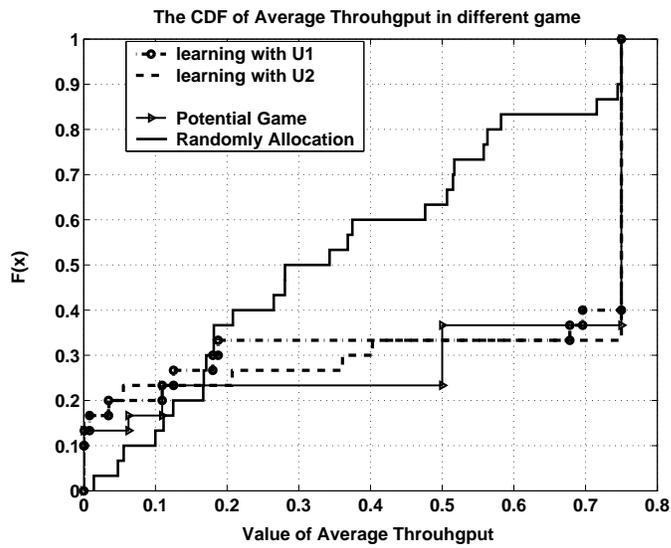}}
\caption{The CDF of Average Throughput} \label{fig:avgcdfthdyspan}
\end{figure}

Figure \ref{fig:avgcdfsirdyspan} shows the CDF of Time Average SIR
in different games.  All learning games and the potential game
outperform the random channel allocation scheme.  The potential
game has the best throughput performance, followed closely by the
cooperative learning scheme. It can be seen in Figure
\ref{fig:avgcdfthdyspan} that half of the users have an average
throughput below 0.3 in the random allocation scheme. The
percentage of users whose average throughput is below 0.3 is 23\%
in potential game, 27\% for learning using $U2$ and 34\% for
learning using $U1$, while the fraction is 51\% for the random
selection.

\begin{figure} [ht]
\centerline{ \epsfxsize=3.5 in\epsffile{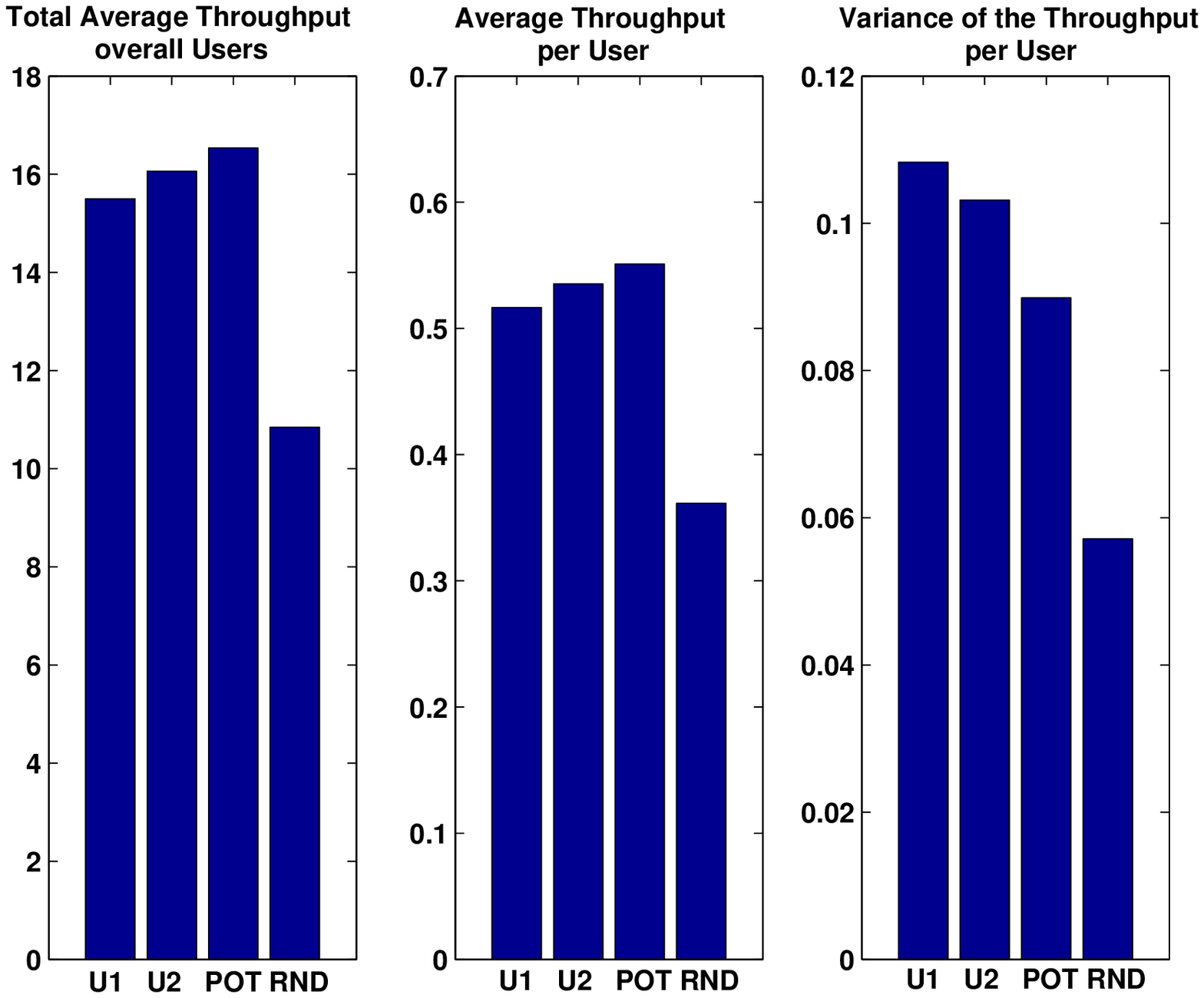}}
\caption{Total Average-Throughput, The Mean and the Variance of the
Throughput per user} \label{fig:avgtotdyspan}
\end{figure}

In Figure \ref{fig:avgtotdyspan} we summarize the performance
comparisons among the proposed schemes: total average throughput,
average throughput per user, and variance of the throughput per
user. The variance performance measure quantifies the fairness,
with the fairest scheme achieving the lowest variance. Among all
the proposed schemes, potential channel allocation game has the
best performance. It is interesting to note that in terms of
average obtained throughput per user, the three schemes perform
very similar, but differ in the performance variability across
users. It seems that even when cooperation is enforced by
appropriately defining the utility, the potential game formulation
provides a fairness advantage over the no-regret learning scheme.


\section{Conclusion}
In this work, we have investigated the design of channel sharing
etiquette for cognitive radio networks for both cooperative and
non-cooperative scenarios. Two different formulations for the
channel allocation game were proposed: potential game formulation,
and no-regret learning. We showed that all the proposed spectrum
sharing policies converge to a channel allocation equilibrium,
although a pure strategy allocation can be achieved only for
cooperative scenarios. Our simulation results have showed that the
average performance in terms of SIR or achievable throughput is
very similar for both learning and potential game formulation, even
for the case of selfish users. However, in terms of fairness, we
showed that both cooperation and allocation strategy play an
important role. While the proposed potential game formulation
yields the best performance, its applicability is limited to
cooperative environments and significant knowledge about
neighboring users is required for the implementation. By contrast,
the proposed no-regret learning algorithm is suitable for
non-cooperative scenarios and requires only a minimal amount of
information exchange.

\linespread{1}
\section{Appendix}

\hspace*{+0.5cm}\textit{Proof:}  Suppose there is a potential
function of game $\Gamma$ :
\begin{equation}
Pot'(S)=\sum^{N}_{i=1}\left( -a \sum^{N}_{j\neq{i},
j=1}p_jG_{ij}f(s_j, s_i) - (1-a) \sum^{N}_{j\neq{i},
j=1}p_iG_{ji}f(s_i, s_j) \right) \label{eq:Potab}
\end{equation}
\hspace*{+0.5cm}where $0<a<1$. Then for all $i \in{\{1, 2, ...,
N\}}$,
\begin{equation}
\hspace*{-1cm}Pot'(s_i, s_{-i})=\sum^{N}_{i=1}\left( - a
\sum^{N}_{j\neq{i}, j=1}p_jG_{ij}f(s_j, s_i) -
(1-a)\sum^{N}_{j\neq{i}, j=1}p_iG_{ji}f(s_i, s_j) \right)\nonumber
\end{equation}
\begin{equation}
=-a\sum^{N}_{j\neq{i}, j=1}p_jG_{ij}f(s_j, s_i) -
(1-a)\sum^{N}_{j\neq{i}, j=1}p_iG_{ji}f(s_i, s_j)\nonumber
\end{equation}
\begin{equation}
+\sum^{N}_{k\neq{i},  k=1}\left[ - a \sum^{N}_{j\neq{k},
j=1}p_jG_{kj}f(s_j, s_k) - (1-a)\sum^{N}_{j\neq{k},
j=1}p_kG_{jk}f(s_k, s_j) \right]\nonumber
\end{equation}
\begin{equation}
= - a \sum^{N}_{j\neq{i} j=1}p_jG_{ij}f(s_j, s_i) -
(1-a)\sum^{N}_{j\neq{i} j=1}p_iG_{ji}f(s_i, s_j)\nonumber
\end{equation}
\begin{equation}
+ \sum^{N}_{k\neq{i},  k=1}\left[- ap_iG_{ki}f(s_i, s_k)- a
\sum^{N}_{j\neq{k},  j\neq{i} j=1}p_jG_{kj}f(s_j,
s_k)\right.\nonumber
\end{equation}
\begin{equation}
\left.-(1-a)p_kG_{ik}f(s_k, s_i) - (1-a)\sum^{N}_{j\neq{k},
j\neq{i},j=1}p_kG_{jk}f(s_k, s_j) \right]\nonumber
\end{equation}
\begin{equation}
=- a \sum^{N}_{j\neq{i},  j=1}p_jG_{ij}f(s_j, s_i) -
(1-a)\sum^{N}_{j\neq{i},  j=1}p_iG_{ji}f(s_i, s_j)\nonumber
\end{equation}
\begin{equation}
+\sum^{N}_{k\neq{i},  k=1}\left(- ap_iG_{ki}f(s_i, s_k)\right)
+\sum^{N}_{k\neq{i}, k=1}\left(- (1-a)p_kG_{ik}f(s_k,
s_i)\right)\nonumber
\end{equation}
\begin{equation}
+\sum^{N}_{k\neq{i},  k=1}\left(- a \sum^{N}_{j\neq{k}, j\neq{i},
j=1}p_jG_{kj}f(s_j, s_k)  - (1-a)\sum^{N}_{j\neq{k},
j\neq{i},j=1}p_kG_{jk}f(s_k, s_j) \right)\nonumber
\end{equation}
\begin{equation}
=- a \sum^{N}_{j\neq{i},  j=1}p_jG_{ij}f(s_j, s_i) -
(1-a)\sum^{N}_{j\neq{i},  j=1}p_iG_{ji}f(s_i, s_j)\nonumber
\end{equation}
\begin{equation}- a\sum^{N}_{k\neq{i},  k=1}p_iG_{ki}f(s_i, s_k)\
-(1-a)\sum^{N}_{k\neq{i},  k=1}p_kG_{ik}f(s_k, s_i) \nonumber
\end{equation}
\begin{equation}
+\sum^{N}_{k\neq{i},  k=1}\left(- a \sum^{N}_{j\neq{k},  j\neq{i}
j=1}p_jG_{kj}f(s_j, s_k) -(1-a)\sum^{N}_{j\neq{k},  j\neq{i}
j=1}p_kG_{jk}f(s_k, s_j) \right)\nonumber
\end{equation}
Let
\begin{equation}
 Q(s_{-i})=\sum^{N}_{k\neq{i}  k=1}\left(- a
\sum^{N}_{j\neq{k}  j\neq{i}  j=1}p_jG_{kj}f(s_j, s_k)  -
(1-a)\sum^{N}_{j\neq{k} j\neq{i}   j=1}p_kG_{jk}f(s_k, s_j)
\right),\nonumber
\end{equation}
Then,
\begin{equation}
Pot'(s_i, s_{-i})=- a \sum^{N}_{j\neq{i},  j=1}p_jG_{ij}f(s_j,
s_i) - (1-a) \sum^{N}_{j\neq{i},  j=1}p_iG_{ji}f(s_i,
s_j)\nonumber
\end{equation}
\begin{equation}
- a\sum^{N}_{k\neq{i},  k=1}p_iG_{ki}f(s_i, s_k)
-(1-a)\sum^{N}_{k\neq{i},  k=1}p_kG_{ik}f(s_k,
s_i)+Q(s_{-i})\nonumber
\end{equation}
\begin{equation}
= - (a+(1-a)) \sum^{N}_{j\neq{i},  j=1}p_jG_{ij}f(s_j, s_i) -
(a+(1-a))\sum^{N}_{j\neq{i},  j=1}p_iG_{ji}f(s_i,
s_j)+Q(s_{-i})\nonumber
\end{equation}
If user $i$ changes its strategy from $s_i$ to $s'_i$, we can get:\\
\begin{equation}
Pot'(s'_i, s_{-i})=- a \sum^{N}_{j\neq{i},  j=1}p_jG_{ij}f(s_j,
s'_i) - (1-a)\sum^{N}_{j\neq{i},  j=1}p_iG_{ji}f(s'_i,
s_j)\nonumber
\end{equation}
\begin{equation}
- a\sum^{N}_{k\neq{i},  k=1}p_iG_{ki}f(s'_i, s_k)
-(1-a)\sum^{N}_{k\neq{i},  k=1}p_kG_{ik}f(s_k,
s'_i)+Q(s_{-i})\nonumber
\end{equation}
\begin{equation}
= - (a+(1-a)) \sum^{N}_{j\neq{i},  j=1}p_jG_{ij}f(s_j, s'_i) -
(a+(1-a))\sum^{N}_{j\neq{i},  j=1}p_iG_{ji}f(s'_i,
s_j)+Q(s_{-i})\nonumber
\end{equation}
Here $Q(s_{-i})$ is not affected by the strategy changing of user
$i$. Hence,
\begin{equation}
Pot'(s'_i, s_{-i}) - Pot'(s_i, s_{-i})= - (a+(1-a))
\sum^{N}_{j\neq{i},  j=1}p_jG_{ij}f(s_j, s'_i)\nonumber
\end{equation}
\begin{equation}
 -(a+(1-a))\sum^{N}_{j\neq{i},  j=1}p_iG_{ji}f(s'_i, s_j)\nonumber
\end{equation}
\begin{equation}
-\left(-(a+(1-a)) \sum^{N}_{j\neq{i}, j=1}p_jG_{ij}f(s_j, s_i)
-(a+(1-a))\sum^{N}_{j\neq{i},  j=1}p_iG_{ji}f(s_i,
s_j)\right)\nonumber
\end{equation}
\begin{equation}
= - \sum^{N}_{j\neq{i}, j=1}p_jG_{ij}f(s_j, s'_i) -
\sum^{N}_{j\neq{i} j=1}p_iG_{ji}f(s'_i, s_j)-\left( -
\sum^{N}_{j\neq{i} j=1}p_jG_{ij}f(s_j, s_i) - \sum^{N}_{j\neq{i},
j=1}p_iG_{ji}f(s_i, s_j)\right)\nonumber
\end{equation}

From equation (\ref{eq:U2}),
\begin{equation}
U_i(s'_i, s_{-i})-U_i(s_i, s_{-i})=-\sum^{N}_{j\neq{i}
j=1}p_jG_{ij}f(s_j, s'_i) -\sum^{N}_{j\neq{i} j=1}p_iG_{ji}f(s'_i,
s_j)\nonumber
\end{equation}
\begin{equation}
- \left( -\sum^{N}_{j\neq{i}  j=1}p_jG_{ij}f(s_j, s_i)
-\sum^{N}_{j\neq{i} j=1}p_iG_{ji}f(s_i, s_j) \right) \forall i=1,
2, ..., N, \nonumber
\end{equation}

\begin{equation}
U_i(s'_i, s_{-i})-U_i(s_i, s_{-i})=Pot'(s'_i, s_{-i}) - Pot'(s_i,
s_{-i}) \forall i=1, 2, ..., N,\nonumber
\end{equation}
So, $Pot'(S)$ in (\ref{eq:Potab}) is an exact potential function
of game $\Gamma$. If we set $a$ to $\frac{1}{2}$ in
(\ref{eq:Potab}), $Pot'(S)$ is the same as $Pot(S)$ defined in
(\ref{eq:Pot}), and we prove that (\ref{eq:Pot}) is an exact
potential function of game $\Gamma$.

\nocite{} \pagebreak


\begin{thebibliography}{1}

\bibitem{Mitola 2000}
J. Mitola III, "Cognitive Radio: An Integrated Agent Architecture
for Software Defined Radio" Doctor of Technology Dissertation,
Royal Institute of Technology (KTH), Sweden, May, 2000

\bibitem{FCC 2005}
"Facilitating Opportunities for Flexible, Efficient, and Reliable
Spectrum Use Employing Cognitive Radio Technologies" FCC Report and
Order, FCC-05-57A1, March 11, 2005

\bibitem{Mitola 1999}
J. Mitola III, "Cognitive Radio for Flexible Mobile Multimedia
Communications",  IEEE 1999 Mobile Multimedia Conference (MoMuC,
November, 1999).

\bibitem{Chakravarthy Shaw 2005}
 V.D. Chakravarthy, A.K. Shaw, M.A. Temple, J.P. Stephens, "Cognitive
radio - an adaptive waveform with spectral sharing capability"
Wireless Communications and Networking Conference, 2005 IEEE Volume
2,  13-17 March 2005 Page(s):724 - 729

\bibitem{Landsford 2004}
J. Lansford, "UWB coexistence and cognitive radio"  Ultra Wideband
Systems, 2004. Joint UWBST and IWUWBS. 2004 International Workshop
Joint with Conference on Ultrawideband Systems and Technologies. on
18-21 May 2004 Page(s):35 - 39

\bibitem{Yamaguchi 2004}
H. Yamaguchi, "Active interference cancellation technique for
MB-OFDM cognitive radio"  Microwave Conference, 2004. 34th European
Volume 2,  13 Oct. 2004 Page(s):1105 - 1108

\bibitem{Goodman_Mandayam 2001}
David J. Goodman and Narayan B. Mandayam, ¡°Network Assisted Power
Control for Wireless Data¡±, Mobile Networks and Applications, vol.
6, No. 5, pp. 409- 415, 2001
%
%\bibitem{Feng_Mau_Mandayam 2004}
%N. Feng and S-C. Mau and N. Mandayam, ¡°Pricing and power control
%for joint network-centric and user-centric radio resource
%management¡±, IEEE Transactions on Communications, September 2004.
%
%\bibitem{Ginde_Neel_Buehrer 2003}
%S. Ginde, J. Neel, and R. Buehrer. "Game Theoretic Analysis of
%Joint Link Adaptation and Distributed Power Control in GPRS",
%Vehicular Technology Conference, Orlando October 2003.
%
%\bibitem{Krishnaswamy 2002}
%D. Krishnaswamy, "Game-theoretic formulations for network-assisted
%resource management in wireless networks" , Proc. of IEEE Vehicular
%Technology Conference, Vancouver, September 2002.
%
%\bibitem{Wong_Wassell 2002}
%Shin Horng Wong, Ian J. Wassell, Application of Game Theory for
%Distributed Dynamic Channel Allocation . IEEE 55th Vehicular
%Technology Conference, Spring 2002, Birmingham, AL, Pages 404-408,
%May 2002.

\bibitem{Menon_Reed 2004}
R. Menon, A. MacKenzie, R. Buehrer, J. Reed, "Game Theory and
Interference Avoidance in Decentralized Networks" SDR Forum
Technical Conference November 15-18, 2004.

\bibitem{Neel_Reed_Gilles 2002}
J. Neel, J.H. Reed, R.P. Gilles " The Role of Game Theory in the
Analysis of Software Radio Networks", SDR Forum Technical
Conference November, 2002.

\bibitem{Neel_Reed_Gilles 2004}
J. Neel, J.H. Reed, R.P. Gilles. "Convergence of Cognitive Radio
Networks," Wireless Communications and Networking Conference 2004.

\bibitem{Hasan}
H. Mahmood,"Investigation of Low Rate Channel Codes for
Asynchronous DS-CDMA", M.Sc Thesis, University of Ulm, Ulm,
Germany, August 2002.

\bibitem{Monder 1996}
D. Monderer and L. Shapley  ``Potential Games''. \textit{ Games and
Economic Behavior 14},  pp124-143, 1996.

\bibitem{Farago 2002}
J. Farago, A. Greenwald and K. Hall, ``Fair and Efficient Solutions
to the Santa Fe Bar Problem ,'' \textit{ In Proceedings of the
Grace Hopper Celebration of Women in Computing 2002 }. Vancouver,
October, 2002.

\bibitem{Jafari 2001}
A. Jafari, A. Greenwald, D. Gondek and G. Ercal, ¡°On No-Regret
Learning, Fictitious Play, and Nash Equilibrium ¡±, In Proceedings
of the Eighteenth International Conference on Machine Learning,
pages 226-223, Williamstown, June, 2001.

\bibitem{Greenwald 2003}
A. Greenwald, A. Jafari, ``A Class of No-Regret Algorithms and
Game-Theoretic Equilibria '' \textit{Proceedings of the 2003
Computational Learning Theory Conference. Pages 1-11, August, 2003.
}

\bibitem{Freund 1995}
Y. Freund and R. Schapire, ``A decision-theoretic generalization of
on-line learning and an application to boosting'', \textit{In
Computational Learning Theory: Proceedings of the Second European
Conference, pages 23-37,} Springer-Verlag, 1995


\end{thebibliography}
\end{document}